\documentclass[pra,twocolumn,superscriptaddress,showpacs,nofootinbib]{revtex4}

\newcommand{\ket}[1]{| #1 \rangle}
\newcommand{\bra}[1]{\langle #1 |}

\usepackage{amsmath}
\usepackage{amsfonts}
\usepackage{amssymb}
\usepackage{bm}
\usepackage{latexsym}

\begin{document}

\title{Information-theoretic temporal Bell inequality and quantum computation}

\author{Fumiaki Morikoshi}
 \email{fumiaki@will.brl.ntt.co.jp}
 \affiliation{NTT Basic Research Laboratories, NTT Corporation, \\
 3-1 Morinosato-Wakamiya, Atsugi, Kanagawa, 243-0198 Japan}
 \affiliation{PRESTO, JST, 4-1-8 Honcho, Kawaguchi, Saitama, 332-0012 Japan}

\begin{abstract}

An information-theoretic temporal Bell inequality is formulated to contrast
classical and quantum computations.
Any classical algorithm satisfies the inequality, while quantum ones can
violate it.
Therefore, the violation of the inequality is an immediate consequence of
the quantumness in the computation.
Furthermore, this approach suggests a notion of {\it temporal nonlocality} in
quantum computation.

\end{abstract}


\maketitle

\section{\label{Sec:Intro}Introduction}

Separation between classical and quantum phenomena plays a crucial role in
understanding the weirdness of quantum physics and has recently been
extensively investigated, especially in the context of quantum information
science. 
One way of drawing a rigid distinction between them is to show the violation of
the Bell inequality \cite{Bell1964}.
In this paper, we develop a Bell-type inequality that distinguishes between
classical and quantum computations.
Here, in order to allow arguments based on conditional entropy between
measurement outcomes of two different observables, we employ the
information-theoretic Bell inequalities formulated by Braunstein and Caves
\cite{Braunstein1988,Braunstein1990}.
Furthermore, instead of discussing correlations between observables of
distantly located physical systems, we focus on one and the same physical
system and analyze correlations between measurement outcomes at
{\it different times} as in the argument on the temporal Bell inequality
initiated by Leggett and Garg \cite{Leggett1985}.

The information-theoretic temporal Bell inequality is formulated for classical
algorithms.
It is satisfied by any classical algorithm but can be violated by quantum
ones.
As an example to show the violation of the inequality, we discuss the so-called
``database'' search problem, which is formally stated as follows:
Given a black box (oracle) that calculates the function $F$ of
$x \in \{0, \ldots, 2^n -1 \}$ such that, for unknown $s$,
\begin{equation*}
F(x) = \begin{cases}
         0, & x \neq s \\
         1, & x=s,
        \end{cases}
\end{equation*}
find the unknown number $s$ by asking the oracle as few times as possible.
Note that we do not assume any structure within the black box, and thus all we
can do is just input an $n$-bit number and obtain an output.
(Tricks such as requesting some portion of $n$ bits, e.g., whether the first
bit of the $n$ bits is 0 or 1, are not allowed in the present setting.)
It is well-known that it takes $O(2^n)$ oracle queries in classical algorithms
to solve the problem, while it requires only $O(\sqrt{2^n})$ queries
in Grover's algorithm for quantum computation \cite{Grover1997}.
It is shown that the information-theoretic temporal Bell inequality is violated
in Grover's algorithm.

Bell-type inequalities are conventionally invoked to find out whether some
profound physical principle such as local realism \cite{Bell1964} or
macroscopic realism \cite{Leggett1985} holds in quantum theory.
However, the information-theoretic temporal Bell inequality is proposed as just
one way of discriminating between classical and quantum computations.
It is not intended here to try to refute nor confirm any physical principle.

Another motivation behind this work is to advance our understanding of the role
of temporal correlations in quantum information processing.
Temporal correlations in quantum phenomena have been studied in the foundations of quantum theory from the viewpoint of Bell-type inequalities since the
original proposal of the Leggett-Garg inequality \cite{Leggett1985, Paz1993}.
In the context of information processing, it was rephrased quite recently from
a slightly different perspective \cite{Taylor2004,Brukner2004}.
The present approach is developed in the belief that, in some cases, the power
of quantum information processing may also lie in temporal correlations of a
truly quantum nature as well as in spatial correlations due to entanglement.

An information-theoretic aspect of the ``database'' search problem has already
been studied in \cite{Koashi1998}, where the maximum amount of information
obtainable from a single oracle query is analyzed.
The present scenario, on the other hand, mainly focuses on the entire process
of the algorithms and thus rather discusses collective behavior of many
successive oracle queries in terms of information-theoretic quantities.

This paper is organized as follows.
First, the information-theoretic Bell inequalities derived in
\cite{Braunstein1988} are briefly revisited in Sec.~\ref{Sec:Revisited}.
Next, we move to the formulation of an information-theoretic temporal Bell
inequality in Sec.~\ref{Sec:Inequality}.
Then, Sec.~\ref{Sec:Violation} gives a proof of the violation of the
inequality in quantum computation.
In Sec.~\ref{Sec:Discussion}, we discuss some implications of the
inequality, especially a notion of {\it temporal nonlocality} suggested by the
violation of the inequality,  and finally summarize the paper.

\section{\label{Sec:Revisited}Information-theoretic Bell inequalities
revisited}

From a mathematical point of view, the existence of a {\it valid} joint
probability distribution of the measurement outcomes of all relevant
observables is essential to Bell inequalities\footnote{
From a physical point of view, needless to say, it is essential for
the conventional (spatial) Bell inequalities that we be able to draw the
conclusion that quantum theory cannot be described by any local hidden variable
theory.
};
any classical theory can assign definite values to all observables irrespective
of measurement processes in principle and any marginal probability distribution
must be consistent with the entire joint probability distribution, while this
is not generally the case in quantum theory.
The information-theoretic version of Bell inequalities is a paraphrase of this
fact in terms of Shannon entropy \cite{Shannon1948}, which is defined as
$H(X) = - \sum_{x} p(x) \log_2 p(x)$ for an observable $X$ with $\{p(x) \}$
being the probability distribution of measurement outcomes $x$.
Since various inequalities between entropies in information theory are derived
for valid joint probability distributions, the inequalities might not be
satisfied for {\it invalid} joint probability distributions generated in
quantum theory.
This is the crux of the information-theoretic Bell inequalities.
In fact, they are derived by applying  basic inequalities in information theory
to valid joint probability distributions in classical theory as shown below.

Consider, say, four observables $A_0, A_1, A_2, A_3$ and their joint entropy
$H(A_0, A_1, A_2, A_3)$.
The chain rule for conditional entropies yields
$H(A_0, A_1, A_2, A_3) = H(A_3|A_0, A_1, A_2) + H(A_2|A_0, A_1) + H(A_1|A_0)
 + H(A_0)$.
(For details of basic inequalities in information theory, see \cite{Cover1991}, for example.)
Since conditioning never increases entropy, e.g.,
$H(A_2|A_0, A_1) \le H(A_2|A_1)$, we have
\begin{eqnarray}
 \label{Eq:ITfour}
  H(A_0, A_1, A_2, A_3) &\le& H(A_3|A_2) + H(A_2|A_1) \nonumber \\
                          & & {}+ H(A_1|A_0) + H(A_0).
\end{eqnarray}

Our information-theoretic temporal Bell inequality is based on this form, but
before moving on to it, it is worthwhile mentioning how Eq.~(\ref{Eq:ITfour})
is related to the well-known argument on the (spatial) Bell inequality between
two distantly located parties (Alice and Bob) with two qubits \cite{CHSH1969}.
Suppose $A_0$ and $A_2$ (with even subscripts) represent Alice's observables
and $A_1$ and $A_3$ (with odd subscripts) Bob's.
The information contained in some quantities is never smaller than that in a	
subset of them, i.e., $H(A_0, A_3) \le H(A_0, A_1, A_2, A_3)$, thus we have
\begin{equation}
 \label{Eq:spatial}
  H(A_3 | A_0) \le H(A_3|A_2) + H(A_2|A_1) + H(A_1|A_0),
\end{equation}
which is one of the information theoretic (spatial) Bell inequalities given in
\cite{Braunstein1988}.
The conditional entropy between Alice's and Bob's local observables satisfies
Eq.~(\ref{Eq:spatial}) classically, because there exists a valid joint
probability distribution of all the observables.
However, it can be  proved to be violated in quantum theory, although we will
not go into the details of the spatial case here.
Roughly speaking, the existence of a (possibly invalid) joint probability
distribution over all the observables corresponds to the assumption of
{\it realism}, and the consistency of all the marginal probability
distributions corresponds to the assumption of {\it locality} in the spatial
argument.
[Schumacher also proved inequalities similar to Eq.~(\ref{Eq:spatial}) and
described an intuitive geometric picture of violating the inequality
\cite{Schumacher1991}.
Cerf and Adami explored entropic Bell inequalities by considering entropy Venn
diagrams \cite{Cerf1997}.]

The form of Eq.~(\ref{Eq:ITfour}) is suitable for our argument on a temporal
version and can be easily generalized to the one with $L+1$ observables,
\begin{eqnarray}
 \label{Eq:ITN}
  H(A_0, \ldots , A_{L}) &\le& H(A_{L}|A_{L-1}) + \cdots  + H(A_1|A_0)
                               \nonumber \\
                         & & {} + H(A_0).
\end{eqnarray}
In the next section, we switch to an argument on temporal correlations by
choosing observables $A_i$ as those of one and the same physical system at
different times from $t_0$ to $t_{L}$, which correspond to $L$ different
computational steps in our scenario.
This inequality also should be satisfied when there exists a valid joint
probability distribution for all observables, although it may be violated in
time evolution governed by quantum theory.

\section{\label{Sec:Inequality}Information-theoretic temporal Bell inequality}

Next, we derive an information-theoretic temporal Bell inequality by applying
Eq.~(\ref{Eq:ITN}) to the ``database'' search problem mentioned above.
Consider an algorithm solving the problem deterministically with $L$ oracle
queries.
Suppose there exists a bit (or a qubit) in which the output of the oracle query
is stored, and the observables $A_i \ (i=0, \ldots, L)$ correspond to the
two-valued (0 or 1) measurement on the (qu)bit after the $i$th oracle query.
(For a reason stated below, $A_0$ is defined to have the fixed initial value of
the output register ``0.'')
Hence, at least classically, by measuring all observables $A_i$, the solution
to the problem can be found.
For instance, in the most straightforward classical algorithm, we input the
numbers from zero to $2^n-1$ to the oracle successively and read each output
of the oracle as a measurement outcome of $A_i$.
This process requires at most $L=2^n$ queries and yields the solution with
probability 1. 
The aim of this problem is to find the unique item $x=s$ out of $2^n$ equally
likely numbers;
hence, we eventually obtain $n$ bits of information by solving the problem.
Furthermore, the amount of information obtained by measuring the outputs of the
oracle cannot be less than the amount of information that is supposed to be
obtained by solving the problem itself.
Therefore, we have
\begin{equation}
 \label{Eq:information}
 n \le H(A_0, \ldots , A_{L}).
\end{equation}
If it were possible to infer the solution by observing random variables
containing less information than the solution itself, then it would be a
contradiction.
[Note that $H(A_0, \ldots , A_{L})$ might be smaller than $n$ bits in
{\it unsuccessful} algorithms, but here we concentrate only on deterministic
algorithms solving the problem successfully.]

By combining Eqs.~(\ref{Eq:ITN}) and (\ref{Eq:information}), we obtain the
information-theoretic temporal Bell inequality for the present problem,
\begin{equation}
 \label{Eq:TemporalBell}
 n \le  H(A_{L}|A_{L-1}) + \cdots + H(A_1|A_0).
\end{equation}
[Note that the last term on the right-hand side of Eq.~(\ref{Eq:ITN}),
$H(A_0)$, is omitted because $A_0$ is defined to have the fixed value ``0'' so
that this unwanted term always vanishes.]
From a classical point of view, this inequality means that the sum of
the conditional entropy between successive oracle queries is sufficient to
constitute the amount of information contained in the solution.
[In fact, it can be proven that the inequality in Eq.~(\ref{Eq:TemporalBell})
holds, for instance, in the most straightforward classical algorithm given
above by an explicit calculation.]
In other words, suppose there are many identical copies of a classical
computer running the algorithm separately, but only the outputs of two randomly
chosen successive oracle queries are measured and the relevant conditional
entropy is calculated independently;
for the first computer, only $A_2$ and $A_3$ are measured, and for the second
one, only $A_{10}$ and $A_{11}$, and so on. 
Then, accumulating all the small pieces of ``progress'' achieved in different
computers, we can infer the solution to the problem.
This picture makes it easy to draw an analogy with ordinary experiments on
spatial Bell inequalities:
First, prepare many identical copies of an entangled pair and then measure
randomly chosen observables.
In our scenario, the timing of measuring the outputs of the oracle is randomly
chosen instead in the manner of Leggett and Garg.\footnote{
In the original Leggett-Garg inequality, we randomly choose two different
times, say $t_1$ and $t_3$, out of four possibilities, $t_1, \ldots, t_4$,
however.
}
The point is that outputs of the oracle, i.e., observables $A_i$ have
predetermined values and that there exists a valid joint probability
distribution of them in any classical computation.

\section{\label{Sec:Violation}Violation in quantum computation}

As we have seen in the previous section, the information-theoretic temporal
Bell inequality (\ref{Eq:TemporalBell}) holds in any classical algorithm.
However, this is not the case for quantum computation.
In this section, we show that the inequality is actually violated in Grover's
algorithm \cite{Grover1997}.
For details of Grover's algorithm, see \cite{Nielsen2000}.
Let the initial state of a quantum computer be
$\ket{\psi} = \frac{1}{\sqrt{2^n}} \sum_{j=0}^{2^n-1} \ket{j}_i
\ket{0}_o$,
where the first and second registers denoted as $\ket{\ }_i$  and
$\ket{\ }_o$ represent the input
and the output of the oracle, respectively.
The output register is usually defined as $\frac{1}{\sqrt{2}} (\ket{0}_o
- \ket{1}_o)$ instead of $\ket{0}_o$ so that the phase flip
operation is automatically performed and the number of oracle queries is
reduced by half.
Nevertheless, we stick to $\ket{0}_o$ because we need to measure the
outputs of oracle queries in the middle of the computation.
Hence, the Grover iteration is represented as
$G=(2\ket{\psi}\bra{\psi}-I) O \sigma_z O$,
where the first term represents the ``inversion about average'' operation,
$O$ is the oracle operator defined as
$O \ket{x}_i \ket{y}_o = \ket{x}_i
\ket{y \oplus F(x)}_o$,
and $\sigma_z$ is the Pauli matrix acting only on the output register
$\ket{\ }_o$.
[The second oracle operation in $G$ is inevitable in order to erase
({\it undo}) the previous oracle operation and disentangle the output register
from the input one so that the interference due to the ``inversion about
average'' operation works properly.]
When we measure the outputs of oracle queries (as is explained in the next
paragraph), the output register $\ket{\ }_o$ is measured in the
computational basis $\{ \ket{0}, \ket{1} \}$ right after the first oracle
operation $O$ in the relevant Grover iteration.
Note that this measurement is not included in the original version of Grover's
algorithm.
It is introduced here to carry out the scenario showing the violation of our
inequality.

An ``experiment'' to test our inequality goes as follows, which certainly
parallels  the classical argument.
First, we prepare many identical copies of a quantum computer, and then run
Grover's algorithm in each computer independently.
Then, we randomly choose two computational steps $k$ and $k+1$ out of
$O(\sqrt{2^n})$ possibilities for each computer and measure the output register
$\ket{\ }_o$ only at the $k$th and $(k+1)$th Grover iterations.
(We measure it right after the first oracle operations $O$ in each Grover
iteration $G$.)
After the first measurement on the output register, the state of the input
register jumps into either the state we are looking for, $\ket{s}_i$, or
the $(2^n -1)$ dimensional subspace orthogonal to $\ket{s}_i$, which we
denote as
$\ket{\alpha}_i = \frac{1}{\sqrt{2^n-1}}\sum_{j \neq s} \ket{j}_i$.
In either case, the quantum computer continues the rest of the Grover
iteration, i.e., $(2\ket{\psi} \bra{\psi}-I)O\sigma_z$, and thus the
state becomes a superposition of
$\ket{\alpha}_i$ and $\ket{s}_i$ again.
We carry out these procedures in all the quantum computers independently
with random choices of $k\ge0$.\footnote{
For an exact correspondence to the classical scenario, we should not continue
the Grover iteration when we luckily find the solution $s$ in the first
measurement of the output, i.e., in the $k$th Grover iteration.
The succeeding dynamics should be replaced by a trivial one to make the
value of the output register ``0'' thereafter.
However, it is easily shown that the quantity $H(A_{k+1}|A_k)$ in this scenario
is always smaller than that calculated in the present scenario.
That is, if the inequality is violated in our scenario, then it is violated
in the above exact one as well.
Therefore, we choose the one developed in the text to make the following
argument simpler.
}

By an explicit calculation, for $k \ge 1$, we have
$H(A_{k+1}|A_k) = H(\cos ^2 \theta)$, where $\theta$ is defined as
$\cos \frac{\theta}{2} \equiv \sqrt{1-\frac{1}{2^n}}$ and
$H(x) \equiv -x \log_2 x -(1-x) \log_2 (1-x)$ is the binary entropy.
We also have $H(A_1|A_0) = H(\cos ^2 \frac{\theta}{2})$.
In what follows, we prove that inequality (\ref{Eq:TemporalBell}) with
$L = O(\sqrt{2^n})$ is violated.
First, since $\sin ^2 \theta = \frac{1}{2^{n-2}}-\frac{1}{2^{2n-2}}$, we have
$\sin ^2 \theta < \frac{1}{2^{n-2}} \le \frac{1}{2} \ (n \ge 3)$.
Thus, $H(\cos ^2 \theta) = H(\sin ^2 \theta) < H( \frac{1}{2^{n-2}} )$.
Defining $f(x) \equiv -x \log_2 x$, the inequality $f(x) \ge f(1-x)$ holds for
$0 \le x \le \frac{1}{2}$.
Hence, we have $H( \frac{1}{2^{n-2}} ) = f( \frac{1}{2^{n-2}} ) +
f(1- \frac{1}{2^{n-2}} ) \le 2 f(\frac{1}{2^{n-2}} ) = \frac{n-2}{2^{n-3}}$.
Furthermore, we clearly have $H(\cos ^2 \frac{\theta}{2}) <1$.
By combining these inequalities, we can estimate the right-hand side of
inequality (\ref{Eq:TemporalBell}) for $L=\sqrt{2^n}$ as follows:
\begin{eqnarray}
 \label{Eq:proof}
 & & H(A_{\sqrt{2^n}}|A_{\sqrt{2^n}-1}) + \cdots + H(A_2|A_1) + H(A_1|A_0)
 \nonumber \\
 &=& (\sqrt{2^n} -1)  H(\cos ^2 \theta) + H \left( \cos ^2 \frac{\theta}{2}
     \right)
 \nonumber\\
 &<& (\sqrt{2^n} -1) \frac{n-2}{2^{n-3}} +1 \nonumber \\
 &=& \frac{n-2}{2^{(n/2)-3}}- \frac{n-2}{2^{n}-3}+1.
\end{eqnarray}
Therefore, for large $n$ and $L=O( \sqrt{2^n} )$, we obtain
\begin{equation}
 \label{Eq:QED}
  H(A_{L}|A_{L-1}) + \cdots + H(A_1|A_0) < n.
\end{equation}
This violation of the inequality is  an immediate consequence of the
quantumness in the computation.
Note that the present inequality is defined only for algorithms solving the
problem successfully.
Thus possible violations of the inequality in {\it unsuccessful} classical
algorithms are not necessarily due to the quantum nature of the computation.
It is no wonder that the inequality can be violated by unsuccessful classical
algorithms because they fail in accumulating a sufficient amount of information
to reach the solution.\footnote{
This is reminiscent of the condition in standard spatial Bell inequality
experiments that local measurements performed by two distantly located parties,
Alice and Bob, should be spacelike separated.
If they were allowed to communicate and organize their measurement settings,
they could violate Bell inequalities even within classical theories.}

\section{\label{Sec:Discussion}Discussion}

In this section, we discuss some implications of this approach.
First, we deal with some possible objections to it, and then discuss a notion
of {\it temporal nonlocality} suggested by the violation of our inequality.

There may be an objection that unnecessarily interrupting the quantum algorithm
by measurement obviously precludes the computation and seems pointless.
However, the whole point of this paper is not to provide any useful techniques
to improve quantum algorithms but to offer a novel way of characterizing
quantum computation.
In fact, in our argument, many identical copies of a quantum computer are
required and are not fully exploited.
One copy of the quantum computer would be sufficient to achieve fast quantum
computation.
Nevertheless, this does not necessarily mean that the present scenario is
useless, which can be seen in analogy to the well-known argument on spatial
Bell inequality experiments.
We consume many copies of an entangled pair to demonstrate the nonlocal
character of the pair, while we need only one copy to accomplish quantum
teleportation \cite{Bennett1993} or superdense coding \cite{Bennett1992}.
Similarly, we require many copies of a quantum computer and spoil their
computational processes to demonstrate the quantumness in the computation,
while we need only one copy to accomplish fast quantum computation.
Therefore, in spite of the fact that it seems quite indirect and inefficient,
the present approach over numerous computers seems promising as a means of
fully appreciating the essential difference between classical and quantum
computations from a physical point of view.\footnote{
Incidentally, I believe that quite a few physicists would be interested in the
principle underlying {\it qualitative} differences between classical and
quantum computations, even if there were no computational speed-up in quantum
computation.
}

The number of terms in inequality (\ref{Eq:TemporalBell}) is different in
classical and quantum algorithms;
classically, it has $O(2^n)$ terms, while quantumly, $O(\sqrt{2^n})$ terms.
Thus, it might not be surprising that the inequality is violated in quantum
computation.
The point is, however, whether we can assign definite values to the outputs of
all the oracle queries and there exists a valid joint probability distribution
over all the outputs.
Detecting the violation of the inequality is considered as one way of
clarifying this point.
In a conventional view, we usually think that a quantum computer is in a
superposition and thus that there exists no definite value of registers in each
computational step.
This view underlies the standard interpretation of quantum computation:
parallel computation due to quantum superposition.

However, what if we try to interpret Grover's algorithm in a realistic way with
a (global) hidden variable model?
Every input and output of oracle queries should have a predetermined value in
this view.
Nevertheless, the algorithm still succeeds in finding the solution in
$O(\sqrt{2^n})$ steps.
This seems quite bizarre but is possible once we accept this perspective.
The trick is that the quantum computer somehow selects promising inputs to the
oracle without exhausting all the possibilities.
Strangely, the solution $x= s$ is always included in those promising
$O(\sqrt{2^n})$ inputs.
It looks as if the quantum computer partially knew which one would fail if it
were fed into the oracle.
Classically, it is natural to keep excluding the previously used
inputs until we hit the solution.
However, the quantum computer could refer to the outcomes of the oracle queries
corresponding to {\it previously unused} inputs.
In other words, the quantum algorithm seems to be able to reach the solution
without accumulating a sufficient amount of information from a classical point
of view.
Classically, the solution can be found only by accumulating a sufficient amount
of information step by step through the process of excluding unsuccessful
inputs.
Our inequality (\ref{Eq:TemporalBell}) and its violation in Eq. (\ref{Eq:QED})
offer one way of expressing these facts mathematically.

One possible explanation of the above counterintuitive interpretation is that
the quantum computer can invoke {\it temporal nonlocality} to select promising
inputs without exhausting all possibilities.
It looks as if the queries performed by the quantum computer were connected in
a nonlocal way in time.
Therefore, it is fair to say that the violation of our inequality in quantum
computation represents the temporal nonlocality.
As the violation of conventional (spatial) Bell inequalities implies that
quantum theory is intrinsically {\it nonlocal}, the violation of the
information-theoretic temporal inequality suggests the notion of temporal
nonlocality in quantum information processing.

Spatial correlations due to entanglement between many qubits are considered to
be an origin of the power of quantum computation in the standard view of
quantum parallelism.
However, the above interpretation focuses on {\it temporal} correlations
between different computational steps and succeeds in revealing another weird
aspect of quantum computation.
Parallel computation due to quantum superposition is counterintuitive, but
exploiting seemingly {\it unperformed} computation is far more so.
These two different but not conflicting perspectives are based on spatial and
temporal approaches, respectively.

The information-theoretic temporal Bell inequality is formulated only for the
so-called ``database'' search problem here, but it can be generalized to other
problems in principle.
Suppose the amount of information obtained by solving the problem is $I$ bits.
(This can be easily determined as $n$ bits in the {\it unstructured} database
search problem discussed above, although determining it is generally a
nontrivial task.)
Then, in a similar way to deriving Eq.~(\ref{Eq:information}), we have
\begin{equation}
 \label{Eq:information2}
 I \le H(A_0, \ldots, A_L),
\end{equation}
where $A_0, \ldots, A_L$ are some observables from which the solution to the
problem is inferred .
By combining this inequality with Eq.~(\ref{Eq:ITN}), we have
\begin{equation}
 \label{Eq:TemporalBell2}
 I \le  H(A_{L}|A_{L-1}) + \cdots + H(A_1|A_0).
\end{equation}
[The last term on the right-hand side of Eq.~(\ref{Eq:ITN}), $H(A_0)$,
can be omitted as is done in the derivation of Eq.~(\ref{Eq:TemporalBell}).]
This also should be satisfied by any classical algorithm deterministically
solving the problem and  can be violated by quantum algorithms, if any.
It remains open to apply the inequality to other quantum algorithms and
obtain further insights into essential features of quantum computation.

\section{\label{Sec:Summary}Summary}

An information-theoretic temporal Bell inequality was formulated for the
so-called ``database'' search problem.
Provided that the problem is successfully solved with certainty, any classical
algorithm satisfies the inequality, but Grover's algorithm for quantum
computation violates it.
Separating classical computations from quantum ones by means of a Bell-type
inequality suggests a novel way to contrast them, which is similar to analyzing
nonlocal effects due to entanglement.
Furthermore, a notion of temporal nonlocality was suggested by the violation
of the inequality.
I hope the present approach will eventually contribute to shed new light on
the nature of quantum computation and temporal correlations in quantum
information processing in general.

\begin{acknowledgments}
I am grateful to M. Koashi for helpful discussions.
\end{acknowledgments}

\bibliography{TemporalBellv3}

\end{document}